\documentstyle[11pt,newpasp,epsf,psfig]{article}

\begin{document}

\title{2D stellar kinematics of nuclear bars}
\author{Eric Emsellem}
\affil{Centre de Recherche Astronomique de Lyon, 9 av, Charles Andr\'e, 69561 Saint-Genis Laval Cedex, France}

\author{Daniel Friedli}
\affil{Observatoire de Gen\`eve, 51 Chemin des Maillettes, CH-1290 Sauverny, Switzerland}
\begin{abstract}
We just started an observational program to obtain 2D kinematics of nuclear bars, to
be compared to N body simulations of single and double-barred galaxies.
\end{abstract}


We have started a program to study the kinematics of double-barred
galaxies using the Integral Field Spectrograph {\tt OASIS} (CFHT).
Stellar bidimensional velocity fields obtained with OASIS are compared with N body
+ SPH simulations, to constrain the characteristics of the primary and secondary
bars (e.g. the pattern speed, Friedli \& Martinet 1993). We intend to confirm that the secondary bars observed
photometrically (e.g. Wozniak et al. 1995) are truly decoupled systems.

In this paper, we present preliminary results on two prototypical 
double-barred galaxies: NGC~3504, which exhibits nearly aligned 
primary and secondary bars, and NGC~5850, in which the two bar components
are almost orthogonal to each other.

There is a hint of the presence of an $m=1$ mode in the nuclear bar
of NGC~3504, already observed in AOB images obtained by Combes et al. (PUEO/CFHT).
We also accidentally observed a Supernova (1998cf, IAUC 6914) at $\sim 6\arcsec$. 
from the centre of the galaxy (see Fig.~1). 
In NGC~5850, the OASIS data show that the kinematical minor-axis
of the secondary bar nearly coincides with the photometric minor-axis of the primary.

In these two objects, the zero velocity curve of the nuclear
bar is close to the photometric minor-axis of the {\em primary}
bar. Single-barred galaxy models failed to reproduce the observed
kinematics. Although this seems counter-intuitive,
our OASIS observations are perfectly consistent with N body 
simulations of double-barred galaxies performed by Friedli (see also
Leon et al., these Proceedings).

A larger sample of double-barred galaxies is going to be observed
with OASIS including both the gaseous and stellar kinematics.

\begin{figure}
\centerline{\psfig{figure=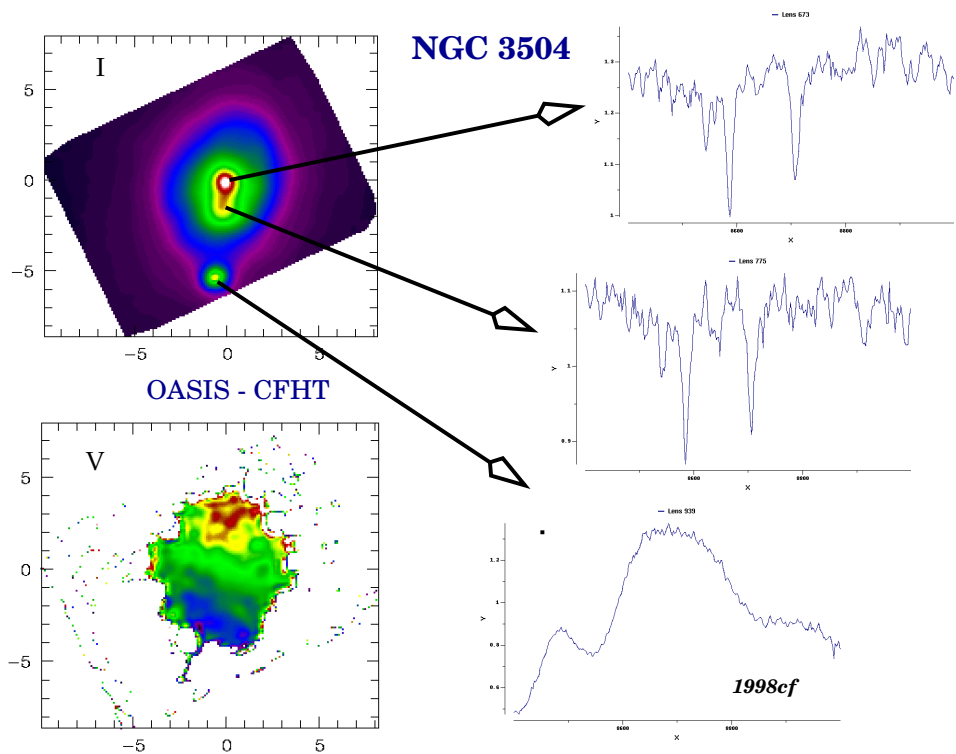,width=13cm,angle=-90}}
\caption{{\tt OASIS} data of NGC~3504: stellar continuum ($I$ band) reconstructed
image (top left), stellar velocity field (bottom left) and three spectra
at different locations on the field of view. The spectrum at the bottom
corresponds to the peak of the detected supernova, and exhibits
broad Ca lines in emission.}
\label{fig:n3504}
\end{figure}

\end{document}